# DESIGN, FABRICATION AND MEASUREMENT OF THE FIRST ROUNDED DAMPED DETUNED ACCELERATOR STRUCTURE (RDDS1)*

J.W. Wang, C. Adolphsen, G.B. Bowden, D.L. Burke, J. Cornuelle, V.A. Dolgashev,
W.B. Fowkes, R.K. Jobe, R.M. Jones, K. Ko, N. Kroll, Z.Li, R.J. Loewen, D. McCormick,
R.H. Miller, C.K. Ng, C. Pearson, T.O. Raubenheimer, R. Reed, M. Ross,
R.D. Ruth, T. Smith, G. Stupakov, SLAC, Stanford, CA, USA
T. Higo, Y. Funahashi, Y. Higashi, T. Higo, N. Hitomi, T. Suzuki, K. Takata, T. Takatomi,
N. Toge, Y. Watanabe, KEK, Tsukuba, JAPAN

*Abstract*

As a joint effort in the JLC/NLC research program, we have developed a new type of damped detuned accelerator structure with optimized round-shaped cavities (RDDS). This paper discusses some important R&D aspects of the first structure in this series (RDDS1). The design aspects covered are the cell design with sub-MHz precision, HOM detuning, coupling and damping technique and wakefield simulation. The fabrication issues covered are ultra-precision cell machining with micron accuracy, assembly and diffusion bonding technologies to satisfactorily meet bookshelf, straightness and cell rotational alignment requirements. The measurements described are the RF properties of single cavities and complete accelerator section, as well as wakefields from the ASSET tests at SLAC. Finally, future improvements are also discussed.

## 1 INTRODUCTION

An active collaboration between KEK and SLAC on the development of X-band accelerating structures began more than 10 years ago. As either joint or parallel efforts, seven JLC/NLC prototypes Detuned (DS) or Damped Detuned Structures (DDS) have been built [1] with several goals in mind. These include testing of long-range wakefield suppression, studying high gradient limits, accelerating beams in NLCTA and improving accelerator efficiency. This last goal has been the main focus behind the latest the Damped Detuned Structure with Round-shaped cavities or RDDS1 (see Figure. 1). Like its predecessors, it is 1.8 m long and consists of 206 cells. To improve the rf efficiency, the structure cells were made with a rounded shape compared to the flat disk shape used in past structures. This change increases the cell shunt impedance by 14%, allowing operation at lower input power to achieve the same average gradient.

## 2 STRUCTURE DESIGN

### 2.1 Cell profile optimisation and accurate dimension calculation

The disc-loaded waweguide cross section for DS, DDS types and RDDS type disks are shown in Figure 2.

*Work supported by U.S. Department of Energy, contract DE-AC03-76F00515 and part of Japan-US Collaboration Program In High Energy Physics Research.

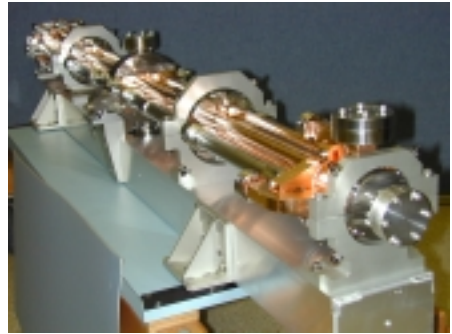

Figure1: RDDS1 section installed on a strongback.

Properly chosen, three arcs in the RDDS design optimized the shunt impedance, Q value and minimized the ratio of peak surface to average accelerating electrical field. Also, the modified HOM coupling slots reduced the leakage of the fundamental mode by a factor of two. Compared to DDS, the overall improvement in shunt impedance for RDDS1 is about 19%.

A newly developed finite element parallel-processing code Omega3P was used for the design calculations [2]. To verify its accuracy, five sets of test disks were carefully measured mechanically with a CMM machine and electrically with a network analyzer. Excellent agreement was obtained between the experimental data and extrapolated quadratic finite element results. A final cell dimensional table was created with a frequency accuracy of better than 1 MHz.

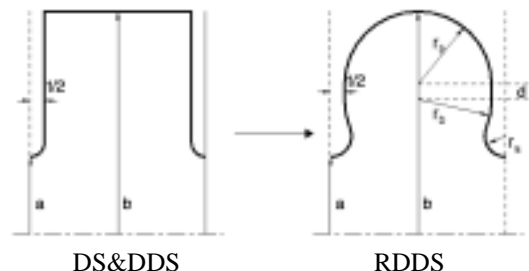

DS&DDS     RDDS

Figure 2: Cell shapes for different disc loaded waveguides.

### 2.2 Design of Accelerator Couplers

The design goal for the input and output fundamental coupler and the high order mode couplers is to provide a good match for the rf signals travelling through each of them. Time domain MAFIA and finite-element OmegaT3 codes were used to simulate these structures, and usually some cold test models were fabricated for confirmation.

Figure 3 shows the cross-sectional view of the input end of RDDS1 structure.

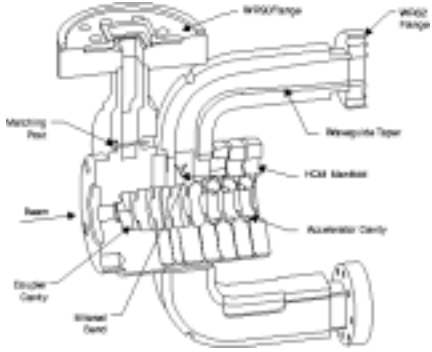

Figure 3: Cutaway view of input end of RDDS1.

## 2.3 Further effort to minimize the long-range transverse wakefield

Space limitations prevented coupling higher order modes to the damping manifolds from the last few cavities. From the ASSET test of an earlier structure (DDS3), it was realised that the pair of fundamental output ports appears to couple to these modes, at least for one polarisation. To exploit this in RDDS1, four symmetric output ports were used that were oriented in the ±45° plane because of space constraints (see Figure 4). Computer simulation and microwave measurements showed that when properly terminated, these last several cavities had Q's as low as 300 [3]. Besides this change in damping, the width of the dipole mode distribution was widened from 10.1 % to 11.3 % to enhance the effect of detuning.

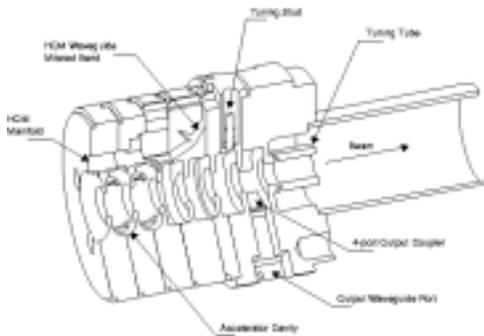

Figure 4: Cutaway view of the output end of RDDS1.

# 3 FABRICATION

## 3.1 Disk Fabrication

After being cut from bar-shaped Class-I OFC copper material, each disk was rough machined to have ~ 40μm remaining copper for all critical dimensions [4]. The final diamond turning was performed on an ultra-high-precision lathe under controlled temperature of 20° ±0.3°C. It was confirmed that the cell profile tolerance of ±1μm was satisfactorily met. A surface finish of 50nm and a flatness better than 0.5μm assured success in stack alignment and diffusion bonding.

## 3.2 Microwave QC and Feed forward

Extensive single and multi-disk rf measurements were made to verify the fundamental and dipole mode frequencies [5]. For single disk QC, each disk was sandwiched between two flat plates. Frequencies deviation of four modes (fundamental zero and π mode, π mode of first dipole band and zero mode of second dipole band) were measured as shown in Figure 5. The deviation of any of the four frequencies is within 0.6 MHz rms.

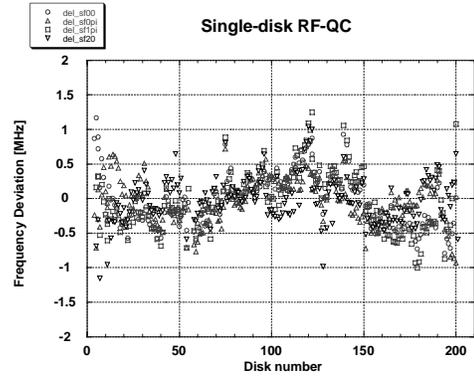

Figure 5: Four frequencies measured in single-disk QC versus cell number.

Systematic errors in the fundamental frequency were compensated by introducing small corrections for subsequent cells. We set the tolerance of 5 degrees in the integrated phase advance at any point along the structure. It should be noted that this criterion is equivalent to a 0.1 MHz systematic frequency error. The beginning disks were machined to the design dimensions and a series of multi-disk microwave measurements were made. Based on this frequency information, a feed-forward correction in "b" (cavity radius) was applied to the disks in later fabrication in order to correct the integrated phase. Figure 6 shows the 2π/3 mode frequencies for six consecutive disks, 2b feed-forward correction and the integrated phase shift (less than 3° finally). With this amount of feed-forward correction, the frequency distribution of the first dipole mode was changed from of the design distribution by less than 1MHz. This will not degrade the designed decoherence for dipole modes [6].

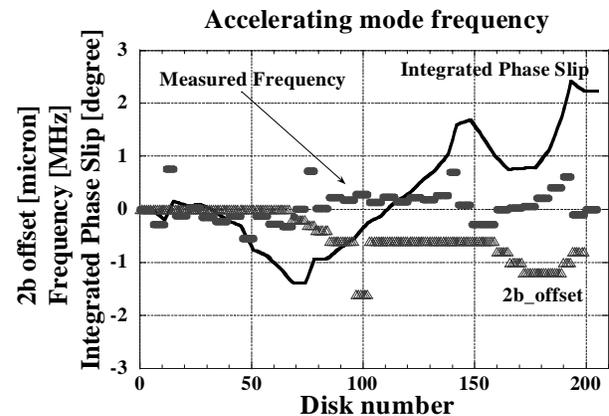

Figure 6: Fundamental mode measurement from stack QC.

*3.3 Structure Assembly*

The disks were manually stacked on a stainless steel V-block, inclined at 60° from horizontal. The disk misalignment was measured by running two capacitive gap sensors along the V-block. An autocollimator was used for monitoring the perpendicularly of the disks. The finished stack was pre-diffusion bonded with 600kg axial force in a furnace at 180°C for a day. The section was then orientated vertically and fully diffusion bonded at 850° - 890°C for four hours. Afterward, waveguide components, cooling and vacuum parts were brazed in a hydrogen furnace. Final measurement showed that the cell-to-cell misalignment was better than ±1 μm and the bookshelving was within 50 μrad. There was a gentle bow of 200μm that was straightened after final assembly on the strongback.

## 4 LESSONS LEARNED

Differential expansion of the endplate supports relative to outer most copper cells lead to a flaring of the ends of the structure during bonding. Likewise, the relative expansion of a stainless-steel torodial vacuum manifold enlarged a few cells near the center of the structure where it was brazed on. Although much of the flaring was removed by symmetrically squeezing the ends of the structure, the schedule for installing the structure in ASSET did not allow time to attempt to correct the center cells (a scheme has since been worked out). Even with the center fundamental frequency errors, which are of the order of 12 MHz, it was decided that much could still be learned from the ASSET measurements.

## 5 ASSET Measurements

The structure was installed in ASSET facility in Sector 2 of the SLAC Linac on May 24 and removed on August 4. [7] As in previous wakefield measurements, positrons extracted from the South Damping Ring served as the drive bunch and electrons extracted from the North Damping Ring served as the witness bunch. Data were taken between PEP II fills, which required streamlining the magnet and timing setup procedure. Parasitic single-beam measurements were also made of the manifold dipole signals versus beam position.

The wakefield results are plotted in Figure 7. Overall there is good agreement between the data and prediction with the frequency errors, which were estimated from bead-pull measurement data of phase shift of the fundamental mode. The wakefield is likely dominated by a few modes in the region of the cell errors. This is supported by the fact that the wakefield phase shows a smooth variation relative to a single frequency oscillation (15.094 GHz).

Additional measurements were taken in a search for higher dipole band contributions in the wakefield dip region where they should be readily apparent. Although

Fnothing obvious was found in this region, 26 GHz wakefields are seen at shorter times once the 15.1 GHz components are removed. This wakefield is likely from modes in the third dipole band. Analysis of higher band contributions is still ongoing.

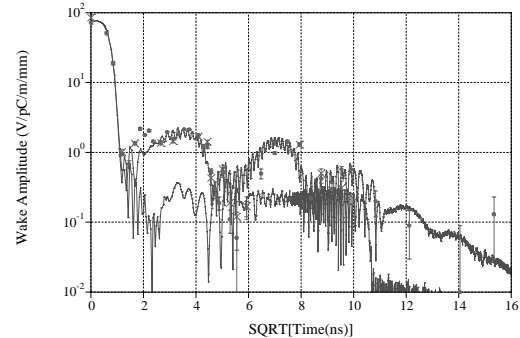

Figure 7: ASSET wakefield data (vertical = dots, horizontal = crosses) and prediction with and without central frequency errors.

For the single beam studies, the manifold dipole signals were either measured with a spectrum analyzer or with dowmmixers and digitizers. In one set of measurements the spectra from the nine RDDS1 ports that had cable connections were recorded. Each measurement was made with the beam roughly centred (based on minimizing the dipole power), and then offset by 0.5 mm in X and Y. The shape of the spectrum with the beam offset in X is in fair agreement with predictions from the same model used to compute the wakefield. However, several narrow monopole-like signals were seen in the data and are not understood. There is a 10% X-Y power coupling of the 14.2 to 14.6 GHz signals that may be related to the cell ellipticity observed near the upstream end of the structure. Finally, the high end of the spectrum falls off smoothly indicating that the fundamental output ports are damping the uncoupled modes as planned.

The wakefield data includes measurements made in both the horizontal and vertical plane: they agree well in both amplitude and phase. For this test, the upstream vertical HOM ports, which couple to the horizontal dipole modes, were blanked off to see if any X-Y differences would be observed. The similarity of the results suggest that the upstream ports have little effect as expected since little power propagates upstream. Therefore, these ports can be removed in future designs.